\documentclass[
    reprint, twocolumn,
    aps,prl,10pt,amsmath,amssymb,
    longbibliography,superscriptaddress,
    showpacs,preprintnumbers,
]{revtex4-2}
\usepackage{xr}
\makeatletter
\newcommand*{\addFileDependency}[1]{
  \typeout{(#1)}
  \@addtofilelist{#1}
  \IfFileExists{#1}{}{\typeout{No file #1.}}
}
\makeatother

\newcommand*{\myexternaldocument}[1]{
    \externaldocument{#1}
    \addFileDependency{#1.tex}
    \addFileDependency{#1.aux}
}

\myexternaldocument{supp}

\usepackage{jmlstyle}
\usepackage{orcidlink}

\definecolor{darkgreen}{rgb}{0,0.5,0}

\begin{document}

\title{Lippmann-Schwinger Approach for Accurate Photoelectron Wavefunctions\\ and Angle-Resolved Photoemission Spectra from First Principles}

\author{Ji Hoon Ryoo}
\email{jhryoo92@gmail.com}
\author{Cheol-Hwan Park\,\orcidlink{0000-0003-1584-6896}}
\email{cheolhwan@snu.ac.kr}
\affiliation{Department of Physics and Astronomy, Seoul National University, Seoul 08826, Korea,}
\affiliation{Center for Theoretical Physics, Seoul National University, Seoul 08826, Korea}

\date{\today}

\begin{abstract}
We present a conceptually simple and technically straightforward method for calculating photoelectron
wavefunctions that is easily integrable with standard wavefunction-based density-functional-theory packages.
Our method is based on the Lippmann-Schwinger equation, naturally incorporating the boundary condition that the final photoelectron state must satisfy.
The calculated results are in good agreement with the measured photon-energy- and polarization-dependence of the angle-resolved photoemission spectroscopy (ARPES) of graphene, the photon-energy-dependent evolution of the so-called dark corridor arising from the pseudospin, and WSe\textsubscript{2}, the circular dichroism reflecting the hidden orbital polarization.
Our study opens doors to do-it-yourself simulations of ARPES with standard density-functional-theory packages, of crucial importance in the era of ``quantum materials,'' whose key experimental tool is ARPES.
\end{abstract}

\maketitle

\paragraph{Introduction.---}{\label{introduction}}

In recent decades, angle-resolved photoemission spectroscopy (ARPES) has become increasingly important following the discovery of materials with fascinating electronic structures, such as topological insulators~\cite{Hsieh2008,Chen2009}, the tunable bandgap of few-layer phosphorene~\cite{Kim2015}, three-dimensional (3D) Weyl semimetals~\cite{Xu2015}, and recently discovered flat-band systems such as twisted bilayer graphene~\cite{Utama2021,Lisi2021} and Kagome superconductor CsV\textsubscript{3}Sb\textsubscript{5}~\cite{hu2022}.

Despite the importance, interpreting ARPES measurements is not always straightforward in many cases.
For example, in the case of graphene, one can completely suppress the photoemission
from a certain Dirac-fermionic branch by tuning the polarization of light~\cite{Gierz2011}. As another example, in
spin-resolved ARPES experiments on a topological insulator
Bi\textsubscript{2}Se\textsubscript{3}, the spin polarization of
photoelectrons depends sensitively on the polarization ~\cite{Jozwiak2013,Ryoo2016}
and energy~\cite{Sanchez-Barriga2014} of light.

Although ARPES simulations based on density-functional theory (DFT)
have achieved success with a wide range of materials, obtaining
information on photoelectrons using standard DFT
packages is often challenging, if possible at all. Since DFT packages
are optimized for supercell geometries, the accurate calculation of the
photoelectron state using DFT has been sparse, except for the ones using
Green-function approaches such as the Korringa--Kohn--Rostoker (KKR)
method~\cite{Ebert2011}.

In this regard, the traditional matrix element calculations mainly
focused on the initial-state effects, approximating the photoelectron
wavefunction \(\left| f \right\rangle\) to be a simple function such as
a plane wave
\(\left| \mathbf{k}_{f} \right\rangle\),
where \(\mathbf{k}_{f}\) denotes the 3D wavevector of the final, photoelectron state. In
fact, in some cases, it has been reported that the calculation based on
the plane-wave approximation yields reasonable results, matching the
measured intensities for a wide range of photoelectron energy values and wavevectors~\cite{Shirley1995,Puschnig2015,Hwang2015,Moser2017}. Within the plane-wave approximation,
the matrix element between the initial and final states, $\left|i\right>$ and $\left|f\right>$, respectively, of the interaction Hamiltonian
\begin{equation}
    \label{eq:Hint}
    H_{\rm int}=\frac{e}{mc}\mathbf{A}\cdot\mathbf{p}\,,
\end{equation}
where $m$ is the electron mass, $e$ the electron charge, $c$ the speed of light, and $\mathbf{p}$ the momentum operator, reads:
\begin{equation}
\left\langle f\left| H_{\rm int} \right|i \right\rangle \approx
\frac{e\hbar}{mc}
\mathbf{A} \cdot \mathbf{k}_{f}\left\langle \mathbf{k}_{f} \middle| i \right\rangle\,.
\end{equation}
Therefore,
if the above approximation holds, then the photoemission intensity is
directly proportional to
\(\left| \left\langle \mathbf{k}_{f}\middle| i \right\rangle \right|^{2}\).
Although this approximation is useful in capturing the initial-state
dependence of the matrix-element effect~\cite{Puschnig2015,Puschnig2009,Dauth2011}, it completely neglects
scattering between photoelectrons and lattice ions. Another serious drawback of the plane-wave
approximation is that it cannot predict the dependence of the photoemission intensity on the light polarization $\mathbf{A}$ other than the trivial geometric factor $\mathbf{A} \cdot \mathbf{k}_f$.
Likewise, this approximation cannot account for the $\mathbf{A}$-dependence of the spin-resolved intensity.
(In passing, we note that the analysis of ARPES at some high-symmetry points can bypass the difficulty of finding the correct final-state wavefunctions~\cite{2017yaji,ryoo2018glide}.)

The method for calculating electronic structure using DFT can be roughly classified into two groups: Green-function methods such as the KKR method~\cite{Ebert2011} and wavefunction-based methods. ARPES simulations have traditionally been studied with the former~\cite{Bansil1984,Henk1993,Ebert2011}, for example, using the SPR-KKR package~\cite{Ebert2011}.
In contrast, although a vast number of researchers use
wavefunction-based DFT packages such as Quantum ESPRESSO~\cite{Giannozzi2009}, VASP~\cite{Kresse1996}, and ABINIT~\cite{Gonze2020},
most of the wavefunction-based DFT software packages are based on periodic boundary conditions; hence, it is conceptually
difficult to represent a photoelectron state, which is inherently
non-periodic in the direction normal to the surface of a solid.
Some wavefunction-based DFT packages, such as SIESTA~\cite{Soler2002} utilize the localized-orbital basis set. In this formalism, one should append a vast number of ``ghost atoms'' to the vacuum to represent a photoelectron state there; unlike valence wavefunctions, wavefunctions of photoelectrons do not decay in the far vacuum.
Therefore, it has been conceptually and technically difficult to simulate ARPES using the standard wavefunction-based DFT packages so far.

Few studies have addressed the calculation of photoelectron wavefunctions by matching wavefunctions across the interface between the bulk material and the surface of a slab supercell~\cite{Krasovskii1997,Krasovskii2004-1,Krasovskii2004-2} or by combining the eigenstates of the slab supercell~\cite{Kobayashi2017,Kobayashi2020,Nozaki2024}. However, these methods may require sophisticated matching algorithms or the computation and selection of several basis functions.
In addition to the fact that only a few groups can simulate ARPES from first principles using the KKR or wavefunction-based methods, these programs are not publicly available. However, in the era of ``quantum materials,'' enabling both experimentalists and theoreticians to easily perform ARPES simulations using standard wavefunction-based DFT packages would profoundly impact the development of the field.

More recently, \citet{DeGiovannini2017JCTC} established a real-time, real-space simulation method based on time-dependent DFT for studying (time-resolved) ARPES, \citet{PhysRevResearch.5.033075} invented an intuitive scattered-wave method, which approximates the final state by a simple extension of the plane wave, and explained the ARPES of graphene, and \citet{Choi2025NatPhys} considered the interference between the Floquet-Bloch states and the Volkov states to explain the pump-probe ARPES experiment on graphene.

In this paper, we present a new wavefunction-based one-step method for
simulating ARPES with accurate photoelectron states within the framework
of DFT. Our new methodology, which uses a modified
version of the Lippmann-Schwinger equation, is conceptually simple and technically easy to implement
in both plane-wave-based and localized-orbital-based DFT software packages. We
demonstrate our method by reproducing the ARPES on graphene, the photon-energy- and polarization-dependence of the so-called dark corridor~\cite{Gierz2012}, and on WSe\textsubscript{2}, the circular dichroism~\cite{Cho2018} revealing the hidden orbital polarization~\cite{Ryoo2017}.

\paragraph{Basic theory of ARPES.---}\label{the-basic-theory-of-arpes}

The intensity of the photoelectron beam can be obtained by Fermi's
golden rule. From first-order time-dependent perturbation theory, the
transition rate from the initial state \(|i\rangle\) to the final state
\(|f\rangle\) by the light-matter interaction reads
\begin{equation}
I \propto \left| \left\langle f\left| H_{\rm int} \right|i \right\rangle \right|^{2}\delta\left( E_{f} - E_{i} - h\nu \right)
\label{eq:Fermi}
\end{equation}
where \(h\nu\) is the photon energy and \(E_{i}\) (\(E_{f}\))
the energy of the initial (final) state. The interaction Hamiltonian [Eq.~\eqref{eq:Hint}] is assumed to be the first-order interaction between a
photon and an electron~\cite{Borstel1985}.

In principle, \(|f\rangle\) can be obtained by solving the Hamiltonian
at a given energy, with the correct boundary condition for
photoelectrons called the time-reversed low-energy electron diffraction
(TLEED) boundary condition~\cite{pendry1976}. For now, we neglect the spin for simplicity.
Suppose that the surface-normal direction is $z$.
We assume that the material
occupies a bounded region $-t_z/2<z<t_z/2$ where $t_z$ is the thickness of the material and is periodic along the in-plane directions. We suppose that the electron detector is at \(z \rightarrow \infty\). Then the asymptotic form of the wavefunction satisfying the TLEED boundary condition~\cite{pendry1976} in vacuum reads
\begin{align}
\label{eq:asympt}
&\psi\left( \mathbf{r}_{\parallel},z\rightarrow\infty \right) = e^{i\mathbf{k} \cdot \mathbf{r}} + \sum_{\mathbf{G}_{\parallel}}{c_{\mathbf{G}_{\parallel}}e^{i\left[(\mathbf{k}_{\mathbf{\parallel}}+\mathbf{G}_{\mathbf{\parallel}}) \cdot \mathbf{r}_{\parallel}- k_{\perp,\mathbf{G}_{\parallel}} z\right]}}\nnnl
&\psi\left( \mathbf{r}_{\parallel},z\rightarrow-\infty \right) =  \sum_{\mathbf{G}_{\parallel}} d_{\mathbf{G}_{\parallel}}\,e^{i\left[(\mathbf{k}_{\mathbf{\parallel}}+\mathbf{G}_{\mathbf{\parallel}}) \cdot \mathbf{r}_{\parallel}+ k_{\perp,\mathbf{G}_{\parallel}} z\right]}\,,
\end{align}
where \(\mathbf{r}_{\parallel} = (x,\,y,\,0)\), ${\bf k}=\mathbf{k}_{\mathbf{\parallel}}+k_z\hat{z}$ with $\mathbf{k}_{\mathbf{\parallel}}\mathbf{=}\left( k_{x}\mathbf{,}\,k_{y}\mathbf{,}\,0 \right)$ the wavevector of the photoelectron propagating toward the detector, i.\,e.\,, $k_z>0$, with energy $E=k^2$ (for the rest of the paper, we adopt the unit where $m=1/2$ and $\hbar=1$), $\mathbf{G}_{\parallel}$ the in-plane reciprocal lattice vector,
\begin{equation}
    \label{eq:kz_Gparallel}
    k_{z,{\bf G}_\parallel}= \sqrt{E-|\mathbf{k}_{\mathbf{\parallel}}+\mathbf{G}_{\mathbf{\parallel}}|^2} \quad\text{ if }|\mathbf{k}_{\mathbf{\parallel}}+\mathbf{G}_{\mathbf{\parallel}}|^2\le E\,,
\end{equation}
and $c_{\mathbf{G}_{\parallel}}$ and $d_{\mathbf{G}_{\parallel}}$ are coefficients.
If $|\mathbf{k}_{\mathbf{\parallel}}+\mathbf{G}_{\mathbf{\parallel}}|^2>E$, $k_{\perp, \mathbf{G}_{\parallel}}$ in Eq.~\eqref{eq:asympt} should be replaced with $-i\kappa_{\perp, \mathbf{G}_{\parallel}}$, where
\begin{equation}
    \label{eq:kappaz_Gparallel}
    \kappa_{z,{\bf G}_\parallel}= \sqrt{|\mathbf{k}_{\mathbf{\parallel}}+\mathbf{G}_{\mathbf{\parallel}}|^2-E} \quad\text{ if }|\mathbf{k}_{\mathbf{\parallel}}+\mathbf{G}_{\mathbf{\parallel}}|^2>E\,,
\end{equation}
so that
the partial wave associated with $\mathbf{G}_{\parallel}$ describes
the electron propagation toward the material from both sides of the vacuum
if \(|\mathbf{k}_{\mathbf{\parallel}}+\mathbf{G}_{\mathbf{\parallel}}|^2\le E\) and the exponential decay of the wavefunction on both sides of the vacuum if \(|\mathbf{k}_{\mathbf{\parallel}}+\mathbf{G}_{\mathbf{\parallel}}|^2> E\),
thus satisfying the TLEED boundary condition: There should be no partial wave associated with $\mathbf{G}_{\parallel}$ propagating away from the material in vacuum~\cite{pendry1976}.

\paragraph{Lippmann-Schwinger
equation.---}\label{the-lippmann-schwinger-equation}

We review the Lippmann-Schwinger equation using a
spinless one-dimensional (1D) Hamiltonian \(H^{\rm 1D} = H^{\rm 1D}_{0} + V(z)\) where
\(H^{\rm 1D}_{0} = p_z^{2} = - d^{2}\text{/}dz^{2}\) and \(V(z)\) is the potential
energy function that is non-zero only in \(-t_z/2<z<t_z/2\). We define the 1D advanced free
Green's function \(G^{\rm 1D}_{0}(E) = \left( E - H^{\rm 1D}_{0} - i\,0^+ \right)^{- 1} \),
the real-space representation of which, $\left\langle z\left| G^{\rm 1D}_{0}(E) \right|z' \right\rangle$, reads, once complete sets of plane-wave integrations $\int dk \left|k\right>\left<k\right|$ are inserted before and after $G^{\rm 1D}_{0}(E)$ and straightforward contour integrations are performed,
\begin{equation}
\left\langle z\left| G^{\rm 1D}_{0}(E) \right|z' \right\rangle =
\begin{cases}
\frac{1}{- 2i\sqrt{E}}e^{- i\sqrt{E}|z - z'|} &\text{if } E \geq 0 \\
\frac{1}{- 2\sqrt{-E}}e^{- \sqrt{-E}|z - z'|} &\text{if } E < 0\,.
\end{cases}
\label{eq:g0_1d}
\end{equation}

The Lippmann-Schwinger equation is given by
\begin{equation}
    \label{eq:LSE}
    \psi = \psi_{0} + G^{\rm 1D}_{0}V\psi\,,
\end{equation}
where \(\psi_{0}\) is an eigenstate of \(H_{0}\) with the same energy
\(E\), i.\,e.\,, \(\psi_{0} = e^{i\sqrt{E}z}\) for the photoelectron heading toward $z\rightarrow\infty$ in the distant future.
From Eq. \eqref{eq:g0_1d},
\begin{equation}
\label{eq:zG0Vpsi_Epositive}
\left< z| G^{\rm 1D}_{0}V\psi \right>\propto
\begin{cases}
e^{- i\sqrt{E}z} &\text{if } E\ge0\text{ and } z>t_z/2 \\
e^{ i\sqrt{E}z} &\text{if } E\ge0\text{ and } z<-t_z/2\,;
\end{cases}
\end{equation}
hence, we find that $\psi$ on the left-hand side of Eq.~\eqref{eq:LSE} automatically satisfies the TLEED boundary condition in this case that there is no $e^{- i\sqrt{E}z}$ component in $\psi$ at $z<-t_z/2$.
The generalization to 3D and spinful systems is straightforward (see End Matter).

\paragraph{Modified Lippmann-Schwinger
equation.---}\label{modified-lippmann-schwinger-equation}

The plane-wave representation works effectively only for periodic
wavefunctions; however, the photoelectron wavefunction is non-periodic along $z$. Since $V({\bf r})$ is
non-zero only for \(-t_z/2 < z < t_z/2,\) we can reconstruct the
wavefunction \(\psi\left( \mathbf{r} \right)\) at any point
\(\mathbf{r}\) once we know the values of
\(\psi\left( \mathbf{r} \right)\) in \(-t_z/2 < z < t_z/2\). To show
this fact, we introduce the cutoff function \(\Theta(z)\), which equals 1 in the material region
\(-t_z/2 < z < t_z/2\) and smoothly decays so that it practically vanishes at the supercell boundaries, \(z = \pm L\text{/}2\). (Note,
however, that if the decay is too fast,
the number of basis functions required for the faithful
representation of the wavefunction will increase.) We define the modified
wavefunction
\begin{equation}
\label{eq:psibar}
\widetilde{\psi}\left( \mathbf{r} \right) = \Theta(z)\psi\left( \mathbf{r} \right).
\end{equation}
By definition,
\(\widetilde{\psi}\left( \mathbf{r} \right) = \psi\left( \mathbf{r} \right)\)
if \(V(\mathbf{r}) \neq 0\). Therefore,
\(V\left( \mathbf{r} \right)\psi\left( \mathbf{r} \right) = V\left( \mathbf{r} \right)\widetilde{\psi}\left( \mathbf{r} \right)\) at any \(\mathbf{r}\). Hence,
using the 3D advanced free Green's function
\(G^{\rm 3D}_{0}(E) = \left( E - H^{\rm 3D}_{0} - i\,0^+ \right)^{- 1} \) with
\(H^{\rm 3D}_{0} = p^{2} = - \nabla^{2}\),
we have
\begin{equation}
\label{eq:LSE2}
\psi = \psi_{0} + G_{0}^{\rm 3D}V\widetilde{\psi}\,,
\end{equation}
which indeed shows that \(\widetilde{\psi}\) contains enough information
to reconstruct \(\psi\) completely. Here, $\psi_0=e^{i\mathbf{k} \cdot \mathbf{r}}\chi$ with $\chi$ the constant spinor of the final state [Eq.~\eqref{eq:psi0}].

We can go one step further and write the Lippmann-Schwinger equation for
\(\widetilde{\psi}\) instead of \(\psi\). Multiplying both sides of
Eq.~\eqref{eq:LSE2} by the cutoff function $\Theta$, we get
\begin{equation}
\left(1-\Theta G^{\rm 3D}_{0}V\right)\widetilde{\psi} = \Theta\psi_{0}\,,
\label{eq:mls}
\end{equation}
from which we solve for $\widetilde{\psi}$;
then, we can obtain the photoelectron state $\psi$ from Eq.~\eqref{eq:LSE2}.
We call Eq.~\eqref{eq:mls} the modified Lippmann-Schwinger equation.
The simple Eqs.~\eqref{eq:mls} and~\eqref{eq:LSE2} are the main result of this study. Moreover, $\psi$ [Eq.~\eqref{eq:LSE2}] is not even necessary for the photoemission matrix element $\left\langle \psi\left| H_{\rm int} \right|i \right\rangle=\left\langle \tilde{\psi}\left| H_{\rm int} \right|i \right\rangle$ because the initial state $\left|i \right\rangle$ is localized at the material, where $\tilde{\psi}=\psi$ [Eq.~\eqref{eq:psibar}].

Our method offers several important advantages: (i) no bulk or slab eigenstates are necessary for obtaining the final, photoelectron state, (ii) no complicated boundary matching process is required because the boundary condition is automatically taken care of by the modified Lippmann-Schwinger equation, and (iii) since all functions in Eq.~\eqref{eq:mls}, \(\widetilde{\psi},\ \Theta\psi_{0}\), and \(V\), are localized near the material and the operator \(\Theta G_{0}^{\rm 3D}\) transforms a localized function into another localized function, Eq.~\eqref{eq:mls} is naturally suitable for standard DFT packages, which adopt periodic boundary conditions. Regarding (iii), we emphasize that our method, which is based on localized periodic functions, is compatible not only with plane-wave-based software packages but also with localized-orbital-based ones, such as SIESTA~\cite{Soler2002}.
In the supercell geometry, localized smooth functions are represented well by a small number of plane
waves or localized orbitals.

A convenient form of the cutoff function \(\Theta(z)\) on the supercell
\(\left\lbrack -L/2,L/2 \right\rbrack\) is given by the
product of the complementary error functions:
\begin{equation}
\Theta(z) = \vartheta\left(-t_z/2-z\right)\vartheta\left(z-t_z/2\right)\,, \\
\end{equation}
where \(\vartheta(z) = \left( 1\text{/}2 \right){\rm erfc}(z\text{/}l_{\rm d})\)
and $l_{\rm d}$ is the length over which the cutoff function decays. A good practice to prevent spurious interactions between periodic images of the material is to set $L$ to the thickness of the material ($t_z$) plus some small vacuum and $l_{\rm d}$ to $10-15\%$ of $L$.

While solving the Schrödinger equation for valence electrons corresponds to solving
an eigenvalue problem, the modified Schrödinger equation [Eq.~\eqref{eq:mls}] is equivalent
to a linear system \(A\widetilde{\psi} = b\) where \(A=1 - \Theta G_{0}V\) is a large square matrix.
This difference arises because the photoelectron states exist at any energy value for a
given in-plane crystal momentum. Unlike the valence-electron problem,
\(A\) is non-Hermitian. In order to solve this large linear system, one
must use iterative methods applicable to non-Hermitian
matrices such as the bi-conjugate gradient stabilized method
(BiCGStab)~\cite{Fletcher1976} or the generalized minimal residual method
(GMRES)~\cite{Saad1986}. We found that BiCGStab achieves convergence with
fewer iterations than GMRES in general.

We implemented the method above to solve photoelectron wavefunctions in
the widely used Quantum ESPRESSO package~\cite{Giannozzi2009}. The lifetimes of both the initial state and the final, photoelectron state were also incorporated. (See End Matter for details.)

\paragraph{Demonstration of the method.---}\label{demonstration-of-the-new-methodology}

\begin{figure}[h!]
\includegraphics[width=0.48\textwidth]{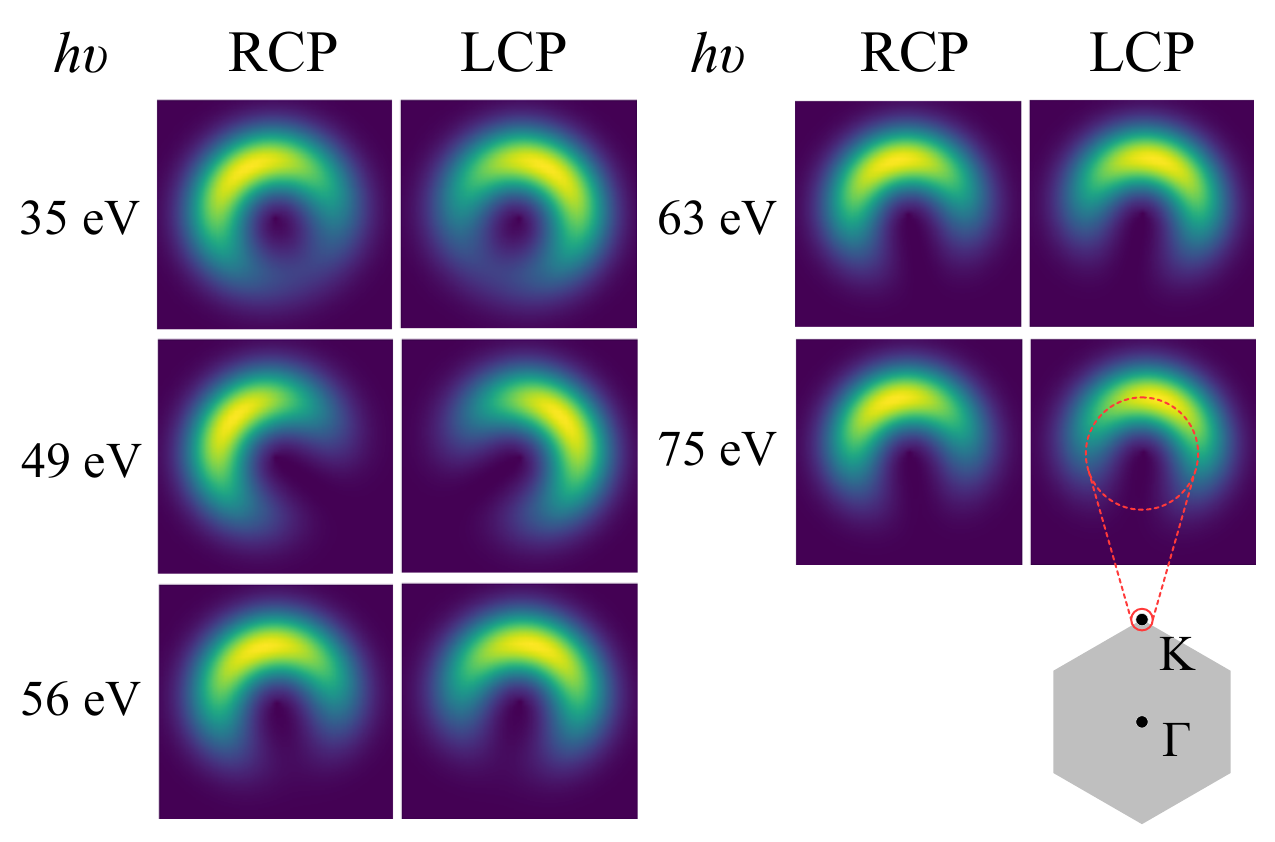}
\caption{ARPES spectra of graphene with photon energy $h\nu$ and the initial state in the upper Dirac band along the contour \(\left| \mathbf{k}_{\parallel} - \mathbf{K} \right| = 0.01\left( 2\pi\text{/}a \right)\), where $a$ is the lattice parameter (see the the Brillouin zone in the lower-right part).
As in Ref.~\onlinecite{Gierz2012}, the \(45^{\circ}\)-incident light lies in the plane parallel to the \(\Gamma\)--K line and the surface normal direction.
}
\label{fig:dark_corridor}
\end{figure}

First, we studied the photon-energy-dependent ``dark corridor'' phenomenon in the ARPES spectra of graphene~\cite{Gierz2012}, in which the ARPES signal from a certain branch of Dirac-like fermions of graphene is completely suppressed depending on the polarization of light. Although the dark corridor is often regarded as a manifestation of the pseudospin degree of freedom of the Dirac-like electronic structure in graphene, the final state significantly affects the exact direction of the dark corridor unless the entire ARPES configuration is symmetric along the mirror plane. Following the experimental setup, we consider \(45^{\circ}\ \)incidence of the right-circularly-polarized (RCP) light and the left-circularly-polarized (LCP) light. We inspected photoemission from the initial states in the upper Dirac band.
Figure~\ref{fig:dark_corridor} shows that the simulated ARPES spectra are in good agreement with the experimental measurements as well as the calculations from the KKR Green-function method, capturing the photon-energy-dependent rotation of the dark corridor for both RCP and LCP photons observed by \citet{Gierz2012} (see Fig.~1 thereof).

We note that our method successfully reproduced the photon-energy dependence of the ARPES of graphene with linearly polarized light also~\cite{PhysRevResearch.5.033075,hong2025}.

\begin{figure}[h!]
\includegraphics[width=0.48\textwidth]{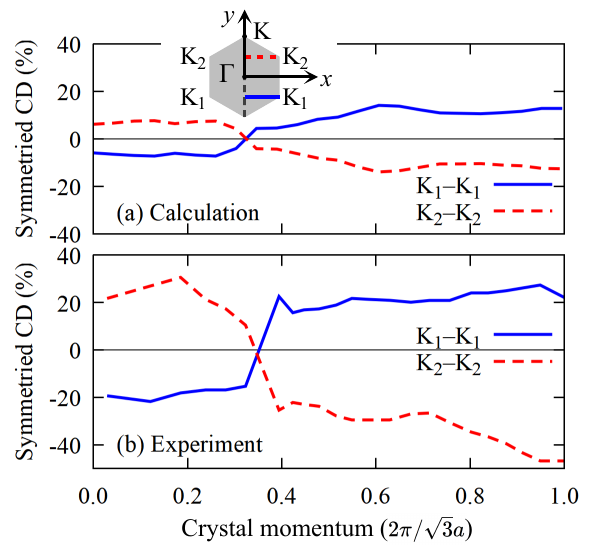}
\caption{(a) The calculated symmetrized CD in the photoemission spectrum from the topmost valence band of bilayer 2H-WSe\textsubscript{2} obtained from our calculations. The symmetrized CD, \(I^{\rm S}_\text{CD}(k_x, k_y)\), is defined as the average of CD at the two k-points connected by the mirror reflection with respect to the \(\Gamma\)--K line so that \(I^{\rm S}_\text{CD}(k_x, k_y) = \frac{1}{2}\left[I_\text{CD}(k_x, k_y) + I_\text{CD}(-k_x, k_y)\right]\) where \(I_\text{CD}(k_x, k_y)\) denotes the CD at a given in-plane momentum \((k_x, k_y)\) of photoelectrons. The in-plane momentum of the initial and photoelectron states are chosen along the path from
\(\frac{2\pi}{a}\left( 0, - \frac{1}{3} \right)\) to
\(\frac{2\pi}{a}\left(\frac{1}{\sqrt{3}}, - \frac{1}{3} \right)\) (solid blue curves) and along the path from \(\frac{2\pi}{a}\left( 0, \frac{1}{3} \right)\) to
\(\frac{2\pi}{a}\left(\frac{1}{\sqrt{3}}, \frac{1}{3} \right)\) (dashed red curves). Here, $a$ is the in-plane lattice parameter, and the corresponding paths in the Brillouin zone are shown in the inset. (b) The experimentally measured symmetrized CD of 2H-WSe$_2$~\cite{Cho2018}.
}
\label{fig:WTe2_ARPES}
\end{figure}

Second, we investigated the non-trivial circular dichroism (CD) in the ARPES spectra of WSe\textsubscript{2}, which is known to exhibit ``hidden'' spin~\cite{Zhang2014} and ``hidden'' orbital~\cite{Ryoo2017} polarizations: spatially localized spin and orbital polarizations at any given crystal momentum in materials having spatial inversion and time-reversal symmetries.
Following the measurement reported by \citet{Cho2018}, we calculated the CD in the photoemission from the topmost valence band of bilayer 2H-WSe\textsubscript{2} using accurate photoelectron wavefunctions obtained fully from first principles for the first time. We used the same ARPES configuration as described in Ref.~\onlinecite{Cho2018} with photon energy set to 94~eV. We chose the initial states along the mirror-symmetry line \(\Gamma\)--K and symmetrized the CD, as done in Ref.~\onlinecite{Cho2018} [see the inset of Fig.~\ref{fig:WTe2_ARPES}(a)].
The dependence of the calculated symmetrized CD on the in-plane wavevector [Fig.~\ref{fig:WTe2_ARPES}(a)] is in good agreement with the measured data in Ref.~\onlinecite{Cho2018} [Fig.~\ref{fig:WTe2_ARPES}(b)]. We emphasize that when it comes to CD, even the best state-of-the-art KKR methods can only achieve qualitative agreement with experimental data~\cite{Gierz2012,Scholz2013,Sanchez-Barriga2014}.

\paragraph{Conclusion.---}\label{conclusion}

We developed a new method for calculating accurate photoelectron wavefunctions within the formalism of DFT.
We proposed solving the Lippmann-Schwinger equation by modifying the equation so that the unknown variable to be solved is the wavefunction multiplied by the cutoff function, making the problem suitable for the supercell geometry with various representations of wavefunctions such as plane-wave or localized-orbital representations. Our method is conceptually very simple, and it does not require the computation of bulk or slab eigenstates or the complex bulk-slab wavefunction-matching process in obtaining the final, photoelectron states. Moreover, the method can be easily implemented in standard wavefunction-based DFT packages. We implemented our methodology in the Quantum ESPRESSO package and applied it to graphene and bilayer WSe\textsubscript{2}. The calculated ARPES spectra show remarkably good agreement with the experimental measurements, successfully capturing the subtle dependence on the energy and polarization of light.

Currently, the computational implementations of reliable and accurate one-step ARPES simulations are accessible to only a few developer groups.
Considering that the majority of the DFT community uses wavefunction-based methods and packages,
the imminent release of our method implemented as an open-source program~\cite{hong2025} for Quantum ESPRESSO~\cite{Giannozzi2009} and our subsequent release of the method implemented as an open-source program~\cite{hong2025} for SIESTA~\cite{Soler2002}, a localized-orbital-based open-source package, will allow the do-it-yourself simulations of ARPES by experimentalists as well as theoreticians, significantly advancing the investigation of quantum materials with fascinating electronic structures.

\begin{acknowledgments}
\paragraph{Acknowledgments.--}
We thank Seung-Ju Hong for sharing the results in Ref.~\onlinecite{hong2025} before publication. This work was supported by the Creative-Pioneering Research Program through Seoul National University and the Korean NRF No-2023R1A2C1007297.
Computational resources have been provided by KISTI (KSC-2023-CRE-0533).
\end{acknowledgments}

\bibliography{main}

\newpage
\onecolumngrid
\vspace{2em}
\begin{center}
\textbf{\large End Matter}
\end{center}
\bigskip
\twocolumngrid

\appendix
\paragraph{3D Lippmann-Schwinger equation.---}

In the 3D case,
\(V({\bf r})\) is non-zero only within $-t_z/2 < z < t_z/2$ and is lattice-periodic in the \(x\) and
\(y\) directions; hence, it is convenient to represent wavefunctions in the
so-called Laue representation, in which we use the real-space
representation in the \(z\) direction and the momentum-space
representation in the \(x\) and \(y\) directions:
\begin{equation}
\phi\left( \mathbf{r} \right) = \sum_{\mathbf{G}_{\parallel}}^{}{\phi_{\mathbf{G}_{\parallel}}(z)e^{i(\mathbf{k}_{\mathbf{\parallel}}+\mathbf{G}_{\mathbf{\parallel}}) \cdot \mathbf{r}_{\parallel}}}\,, \\
\label{eq:Laue}
\end{equation}
where \(\mathbf{r}_{\parallel} = (x,\,y,\,0)\).
In this representation, the application of the 3D
advanced free Green's function reduces to the application of the 1D one for each \(\mathbf{G}_{\mathbf{\parallel}}\) with energy $E-|\mathbf{k}_{\mathbf{\parallel}}+\mathbf{G}_{\mathbf{\parallel}}|^2$, i.\,e.\,,
\begin{align}
    \label{eq:1D_3D_Green}
&G_0^{\rm 3D}(E)\,\phi\left( \mathbf{r} \right) \nnnl
=&\left[E-(p_x^2+p_y^2+p_z^2)-i0^+\right]^{-1}\, \sum_{\mathbf{G}_{\parallel}}{\phi_{\mathbf{G}_{\parallel}}(z)e^{i(\mathbf{k}_{\mathbf{\parallel}}+\mathbf{G}_{\mathbf{\parallel}}) \cdot \mathbf{r}_{\parallel}}}\nnnl
=& \sum_{\mathbf{G}_{\parallel}}\left[E-|\mathbf{k}_{\mathbf{\parallel}}+\mathbf{G}_{\mathbf{\parallel}}|^2-p_z^2-i0^+\right]^{-1}\,{\phi_{\mathbf{G}_{\parallel}}(z)e^{i(\mathbf{k}_{\mathbf{\parallel}}+\mathbf{G}_{\mathbf{\parallel}}) \cdot \mathbf{r}_{\parallel}}}\nnnl
=&\sum_{\mathbf{G}_{\parallel}}{G_0^{\rm 1D}(E-|\mathbf{k}_{\mathbf{\parallel}}+\mathbf{G}_{\mathbf{\parallel}}|^2)\, \phi_{\mathbf{G}_{\parallel}}(z)e^{i(\mathbf{k}_{\mathbf{\parallel}}+\mathbf{G}_{\mathbf{\parallel}}) \cdot \mathbf{r}_{\parallel}}}\,.
\end{align}
The solution of the Lippmann-Schwinger equation with
\(\psi_{0} = e^{i\mathbf{k} \cdot \mathbf{r}}\) is the TLEED state
whose momentum at the detector is ${\bf k}={\bf k}_\parallel+k_z\hat{z}$.
Note that if
\(|\mathbf{k}_{\mathbf{\parallel}}+\mathbf{G}_{\mathbf{\parallel}}|^2>E\) the real-space representation of \(G^{\rm 1D}_{0}(E-|\mathbf{k}_{\mathbf{\parallel}}+\mathbf{G}_{\mathbf{\parallel}}|^2)\) is
proportional to \(e^{- \kappa_{z,{\bf G}_\parallel}|z - z'|}\) [see Eqs.~\eqref{eq:kappaz_Gparallel} and~\eqref{eq:g0_1d}], which may be likened to an evanescent wave.

In the spinful case, in addition to the photoelectron wavevector \(\mathbf{k}\), we also need to
specify the spin direction measured by the detector, characterized by a
constant spinor \(\chi\). The Lippmann-Schwinger equation is unchanged
except that we choose
\begin{equation}
    \label{eq:psi0}
    \psi_{0} = e^{i\mathbf{k} \cdot \mathbf{r}}\chi\,.
\end{equation}

\paragraph{Implementation in DFT software packages.---}

The central equation of our new methodology is the modified
Lippmann-Schwinger equation [Eq. \eqref{eq:mls}]. In this section, we describe
the technical aspects of implementing the solver routine of Eq. \eqref{eq:mls} in
existing DFT software packages.

Since \(G_{0}^{\rm 1D}\)
transforms localized functions (near the material region) into
non-localized ones, we must carefully implement its application.
To accurately and efficiently evaluate \(\left( G_{0}^{\rm 1D}f \right)(z)\) for a given one-dimensional function \(f(z)\) defined on \(\left\lbrack -L/2, L/2 \right\rbrack\), we proceed as follows.
For simplicity of the result, we shift the origin of the $z$ axis to the left
boundary of the simulation region, i.\,e.\,, we assume that
\(f(z)\) is defined on \(\lbrack 0,L\rbrack\). Within this setup, the material is centered at $z=L/2$. The function \(f(z)\)
corresponds to \({\widetilde{\psi}}_{\mathbf{G}_{\parallel}}(z)\), which is the Laue representation [Eq.~\eqref{eq:Laue}] of $\widetilde{\psi}$ [Eq.~\eqref{eq:mls}]. First, we expand \(f(z)\) in the Fourier series:
\begin{equation}
f(z) = \sum_{G_z}^{}{f_{G_{z}}e^{iG_{z}z}}\,,
\end{equation}
where \(G_{z}\) denotes the reciprocal lattice vector along the
surface-normal direction, \emph{z}.
Performing
\(G_{0}^{\rm 1D}\left( E-|\mathbf{k}_{\mathbf{\parallel}}+\mathbf{G}_{\mathbf{\parallel}}|^2 \right)\) on each plane wave,
\(e^{iG_{z}z}\), and using Eq.~\eqref{eq:g0_1d}, we get
\begin{widetext}
\begin{equation}
\label{eq:g01df}
\left( G_{0}^{\rm 1D}f \right)(z) =
\begin{cases}
\frac{1}{- 2ik_{z,{\bf G}_\parallel}}\sum_{G_{z}}f_{G_{z}} \left[
{\frac{e^{iG_{z}z} - e^{- ik_{z,{\bf G}_\parallel}z}}{i\left( G_{z} + k_{z,{\bf G}_\parallel} \right)}}
+ {\frac{e^{- ik_{z,{\bf G}_\parallel}(L - z)} - e^{- iG_{z}(L - z)}}{i\left(G_{z} - k_{z,{\bf G}_\parallel}\right)}}
\right]\quad\text{ if }|\mathbf{k}_{\mathbf{\parallel}}+\mathbf{G}_{\mathbf{\parallel}}|^2\le E\\
\vspace{0.05cm}\\
\frac{1}{- 2\kappa_{z,{\bf G}_\parallel}}\sum_{G_{z}}
f_{G_{z}}  \left[
{\frac{e^{iG_{z}z} - e^{- \kappa_{z,{\bf G}_\parallel}z}}{i G_{z} + \kappa_{z,{\bf G}_\parallel} }}
+ {\frac{e^{- \kappa_{z,{\bf G}_\parallel}(L - z)} - e^{- iG_{z}(L - z)}}{iG_{z} - \kappa_{z}}}
\right]\quad\text{ if }|\mathbf{k}_{\mathbf{\parallel}}+\mathbf{G}_{\mathbf{\parallel}}|^2>E
\end{cases}
\end{equation}
\end{widetext}
for \(0 \leq z \leq L\). (Note that the fractions in the square brackets of Eq.~\eqref{eq:g01df}  remain analytic for vanishing denominators, ensuring numerical
stability.) Therefore,
\(\left( G^{\rm 1D}_{0}f \right)(z)\) can be rewritten as
\begin{widetext}
\begin{equation}
\label{eq:g01df2}
\left( G_{0}^{\rm 1D}f \right)(z) =
\begin{cases}
\alpha e^{- ik_{z,{\bf G}_\parallel}z} + \beta e^{- ik_{z,{\bf G}_\parallel}(L - z)} + \sum_{G_{z}}^{}{\gamma_{G_{z}}e^{iG_{z}z}}\quad\text{ if }|\mathbf{k}_{\mathbf{\parallel}}+\mathbf{G}_{\mathbf{\parallel}}|^2\le E\\
\alpha e^{- \kappa_{z,{\bf G}_\parallel}z} + \beta e^{- \kappa_{z,{\bf G}_\parallel}(L - z)} + \sum_{G_{z}}^{}{\gamma_{G_{z}}e^{iG_{z}z}}\quad\text{ if }|\mathbf{k}_{\mathbf{\parallel}}+\mathbf{G}_{\mathbf{\parallel}}|^2>E
\end{cases}
\end{equation}
\end{widetext}
with appropriate \(\alpha,\ \beta,\) and \(\gamma_{G_{z}}\)'s. Since \(f(z)\) is a
smooth periodic function, \(f_{G_{z}}\) rapidly decreases with
\(|G_{z}|\), and so is \(\gamma_{G_{z}}\), which is proportional to
\(f_{G_{z}}\). Therefore, its real-space representation,
\(\sum_{G_z}^{}{\gamma_{G_{z}}e^{iG_{z}z}}\), can be
efficiently calculated using the fast Fourier transform. Then
\(\alpha e^{-ik_{z,{\bf G}_\parallel}z} + \beta e^{- ik_{z,{\bf G}_\parallel}(L - z)}\)
or
\(\alpha e^{-\kappa_{z,{\bf G}_\parallel}z} + \beta e^{- \kappa_{z,{\bf G}_\parallel}(L - z)}\)
is evaluated in the real-space
grid. After multiplying \(\Theta(z)\),
\(\left( \Theta G_{0}^{(1D)}f \right)(z)\) can be transformed back to
the $G_z$ space if one wishes. We emphasize that, unlike
\(\left( G_{0}^{(1D)}f \right)(z)\) which is a non-periodic function,
\(\left( \Theta G_{0}^{(1D)}f \right)(z)\) is periodic, thus, suitable for a plane-wave representation. [We also note that, unlike
\(\left( G_{0}^{(1D)}f \right)(z)\) which is a delocalized function,
\(\left( \Theta G_{0}^{(1D)}f \right)(z)\) is localized, thus, suitable for a localized-orbital representation as well.]

Here are the steps to apply $(1 - \Theta\, G^{\rm 3D}_0\, V)$ to a function $\phi$: (i) Apply the potential $V$, which has both local and non-local parts, to $\phi$. DFT software packages already provide a way to do this job. (ii) Convert $V\phi$ into Laue representation [Eq.~\eqref{eq:Laue}]. (iii) Calculate $\Theta\,G^{\rm 3D}_0\, (V\phi)$ using Eqs.~\eqref{eq:1D_3D_Green}--\eqref{eq:g01df2}.

The asymptotic time complexity of the algorithm described above is
\(O\left( N_{z}{\log N}_{z} \right)\), where \(N_{z}\) denotes the
number of reciprocal lattice vectors in the \emph{z} direction, i.\,e.\,, $G_z$'s, or the number of the
real space grid along $z$. This is much faster than a naive method for
calculating \(G_{0}^{(1D)}f\) by evaluating the integral for each
real-space grid point, whose time complexity is \(O\left( N_{z}^{2} \right)\). Also, in the naive real-space integration method, the discretization error is
asymptotically polynomial as \(N_{z} \rightarrow \infty\); for example,
the error asymptotically approaches \(O\left( N_{z}^{- 2} \right)\) for numerical integration using the
mid-point rule. On the other hand, the method described above uses the
rapidly decaying Fourier coefficients and thus is far more accurate.

The size of the linear system in Eq.~(12) is determined by the number of basis functions used to represent the wavefunctions and potentials, as in any DFT calculations. For our calculations on graphene and WSe$_2$ with plane-wave basis sets, the typical matrix size was on the order of $10^5 \times 10^5$. An iterative solver avoids constructing the matrix explicitly; instead, it only requires a procedure to compute the action of the operator $(1 - \Theta\, G^{\rm 3D}_0\, V)$ on a given function, which is highly efficient.

On the other hand, our method can also be implemented in software packages based on localized basis sets; in such cases, the construction of the matrix representation of $(1 - \Theta\, G^{\rm 3D}_0\, V)$ and its direct inversion is also an option.

\paragraph{Finite-lifetime effects.---}
\label{finite-lifetime-effects}

In this section, we discuss how our method incorporates many-body effects on the initial and final states in photoemission processes.

Many-body effects on initial states, i.\,e.\,, shift and broadening of the valence bands near the Fermi energy can be systematically calculated using the {\it GW} approximation~\cite{Hybertsen1986} for electron-electron interactions and within the Allen-Heine-Cardona approximation~\cite{2017GiustinoRMP} for electron-phonon interactions. We can incorporate the linewidth of the initial state by replacing the energy-conservation delta function \(\delta(E_f-E_i-h\nu)\) in Eq.~\eqref{eq:Fermi} with the Lorentzian
\begin{equation}
    \label{eq:broadening_initial}
    \frac{1}{\pi}\,\frac{|\Sigma''_i|}{(E_f-E_i-h\nu)^2+\left(\Sigma''_i\right)^2}\,,
\end{equation}
where $\Sigma''_i$ is the imaginary part of the self-energy of the initial state.
On the other hand, we consider the linewidth of the final state in calculating the final-state wavefunction \(|f\rangle\) by using  \(V+\Sigma_{f}\) instead of \(V\), where \(\Sigma_{f}\) is the self-energy of the photoelectron.

\paragraph{Details of the first-principles calculations.---}
\label{details-on-the-first-principles-calculations}

We performed DFT calculations using the Quantum ESPRESSO package~\cite{Giannozzi2009}. We simulated graphene and bilayer WSe$_2$ in the supercell geometry, where the thickness of the vacuum between
adjacent periodic copies of the system is set to
\(15\ \mathring{\mathrm{A}}\). The kinetic-energy cutoff for the wavefunctions is set to 120~Ry.
The exchange-correlation interactions are approximated within the scheme of Perdew, Burke, and Ernzerhof~\cite{Perdew1996}.

We generated norm-conserving pseudopotentials that match the
scattering properties of the all-electron potential from the energy of the valence band minimum to 10 Ry above the vacuum level to describe both the valence electrons and the photoelectron faithfully. This condition
requires two projectors per angular momentum channel.

For the initial-state linewidth, we used the value that gives us the broadening in the measured intensity as a function of the in-plane momentum at a constant initial-state energy, which is 0.2~eV for graphene. To incorporate the lifetime of the final state, we set the imaginary self-energy that gives us the proper photon-energy dependence of the photoemission spectrum, which is 4~eV for both graphene and WSe$_2$.

We solved the linear system in Eq.~\eqref{eq:mls} using BiCGStab~\cite{Fletcher1976} and obtained the converged photoelectron wavefunctions without the use of specialized preconditioners. Typically, it took 50--100 iterations for convergence.

\end{document}